\documentstyle[pre,aps,floats,epsf,amsmath,amssymb,psfig,epsfig]{revtex}

\begin{document}

\draft 
\twocolumn[\hsize\textwidth\columnwidth\hsize\csname@twocolumnfalse%
\endcsname

\title{\bf Single-vehicle data of highway traffic - a statistical analysis}

\author{L. Neubert$^1$, L. Santen$^{1,2}$, A. Schadschneider$^{2}$, M.
  Schreckenberg$^1$}

\address{$^1$Theoretische Physik/FB 10,
  Gerhard-Mercator-Universit\"at Duisburg, D--47048 Duisburg, Germany,\\
  email: {\tt neubert,santen,schreck@traffic.uni-duisburg.de}\\
    $^2$Institut f\"ur Theoretische Physik,
    Universit\"at zu K\"oln,
    D--50937 K\"oln, Germany,
    email: {\tt as,santen@thp.uni-koeln.de}}

\date{\today}

\maketitle


\begin{abstract}
  In the present paper single-vehicle data of highway traffic are
  analyzed in great detail. By using the single-vehicle data directly
  empirical time-headway distributions and speed-distance relations
  can be established. Both quantities yield relevant information about
  the microscopic states. Several fundamental diagrams are also
  presented, which are based on time-averaged quantities and compared
  with earlier empirical investigations. In the remaining part
  time-series analyses of the averaged as well as the single-vehicle
  data are carried out. The results will be used in order to propose
  objective criteria for an identification of the different traffic
  states, e.g. synchronized traffic.
\end{abstract}


]

\section{Introduction}

Experimental and theoretical investigations of traffic flow are focus
of extensive research interest during the past decades
\cite{May,Daganzo,juelich,duis,Helbi}. Various theoretical concepts
(e.g. \cite{Lighthill55,Chandler58,Nagel92,Ito95,Bando94}) have been
developed and numerous empirical observations have been reported
\cite{Koshi83,Hall86,Kerner961,Kerner972,Kerner981}. Despite these
enormous scientific efforts both theoretical concepts as well as
experimental findings are still under debate. In particular the
empirical analysis turns out to be very subtle because the data
strongly depend on several external influences, e.g.  weather
conditions, or the performance of junctions \cite{Hall86}. Therefore,
even certain experimental facts are not well established, although
considerable progress has been made in the past few years. To date,
the following experimental view of highway traffic seems to be of
common knowledge and generally accepted: It has been found that at
least three states with qualitatively different behavior exist, namely
free-flow, stop-and-go and synchronized states
\cite{Koshi83,Kerner961,Kerner972,Kerner981}. In addition to the
existence of qualitative different phases some other interesting
phenomena have been observed, e.g. the spontaneous formation of jams
\cite{Treiterer75}, and hysteresis effects \cite{Hall86}.

This work focuses basically on two points. First of all we present a
{\em direct analysis of single-vehicle} data which leads to a more
detailed characterization of the different microscopic states of
traffic flow, and second we use standard techniques of time-series
analysis in order establish {\em objective} criteria for an
identification of the different states.

A more detailed characterization of the microscopic structure of the
different traffic states should lead to sensitive checks of the
different modeling approaches. In particular the time-headway
distributions and the speed-distance relations, which, to our
knowledge for the first time, have been calculated from the
single-vehicle data, allow for a quantitative comparison with
simulation results of microscopic models \cite{Chowdhury97,s2s,asrev}.
Moreover the objective criteria for an identification of the different
traffic states, developed in the framework of this article, allow for
an unbiased analysis of the experimental data.

The paper is organized as follows. In section \ref{data} we present
some technical details of the measurements as well as of the given
data set. The analysis of single-vehicle data is presented in section
\ref{sec:single}. Explicitly we show results for the time-headway
distribution and the speed-distance relations. These results are
compared to earlier estimates based on data from Japanese highways
\cite{Koshi83,Bando94}. In section \ref{sec:fd} we show the results
for the fundamental diagram. Here we focus on the effect of different
time-intervals for the collection of data and discuss different
methods for the calculation of the stationary fundamental diagram.
Finally the time-series analysis of the single-vehicle data as well as
of the aggregated data are presented in section \ref{sec:time}.

\section{Remarks on the data-collection }
\label{data}

The data set is provided by 12 counting loops all located at the
German highway A1 near Cologne. At this section of the highway a speed
limit of $100\,km/h$ is valid -- theoretically.

%
%
\begin{figure}[ht]
  \begin{center}
    \epsfig{file=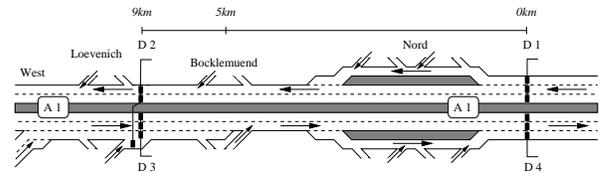,width=0.9\linewidth}
    \caption{Sketch of the analyzed section of the German highway A1. The
      driving directions are given by the arrows. The detector pairs
      D1/D4 and D2/D3 are about $9\,km$ apart, in between a additional
      junction is located.}
    \label{detector}
  \end{center}
\end{figure}
In Fig.~\ref{detector} the section of the highway and the position of
the detectors are sketched. A detector consists of three individual
detection devices, one for each lane. By combining three devices
covering the three lanes belonging to one direction (except D2, see
Fig.~\ref{detector}) one gets the cross-section labeled D1 through D4.
The two detector arrangements D1 and D4 are installed at the
intersection of two highways (AK K\"oln-Nord), while D2 and D3 are
located close to a junction (AS K\"oln-L\"ovenich). These locations
are approximately $9\,km$ apart.  In between there is a further
junction but with a rather low usage.  The most interesting results
are obtained at D1 where the number of lanes is reduced from three to
two for cars passing the intersection towards K\"oln-L\"ovenich.
Therefore, this part of the highway effectively acts as a bottleneck.
Consequently, congested traffic is most often recorded at detector D1,
and the analysis is mainly based on this data set.

The data were collected between June 6, 1996 and June 17, 1996 when a
total number of more than $500,000$ vehicles passed each
cross-section, with a portion of trucks and trailers of about $16\%$
on average. During this period the traffic data set was not biased due
to road constructions or bad weather conditions.

The distance-headway $\Delta x$ as well as the velocity $v$ of the
vehicles passing a detector are collected in the data set. The
velocity $v$ is derived from the time elapsed between the crossing of
the first and the second detector installed in a row with a known
distance (usually $2\,m$). The second direct measure is the time
elapsed between two consecutive vehicles. Due to storage capacity
reasons it is saved with a rough resolution (only $1\sec$), but used
to determine the distance between the vehicle $n+1$ and its
predecessor $n$ via\footnote{This implies that the error in
  calculating $\Delta x_{n+1}$ increases with $\Delta t_{n+1}$, since
  the calculation is made under the assumption of a constant speed
  $v_n$.} $\Delta x_{n+1}=v_n\Delta t_{n+1}$. It should be mentioned
that this procedure gives correct results as far as the velocity {\em
  at} the detector is constant. Here the small spatial extension of
the detectors guarantees that this assumption is fulfilled. So it is
admissible to overcome the restriction of the resolution applying the
reverse procedure to recover $\Delta t$ with higher accuracy as it has
been used in the framework of this paper.

%
%
\begin{figure}[ht]
  \begin{center}
    \epsfig{file=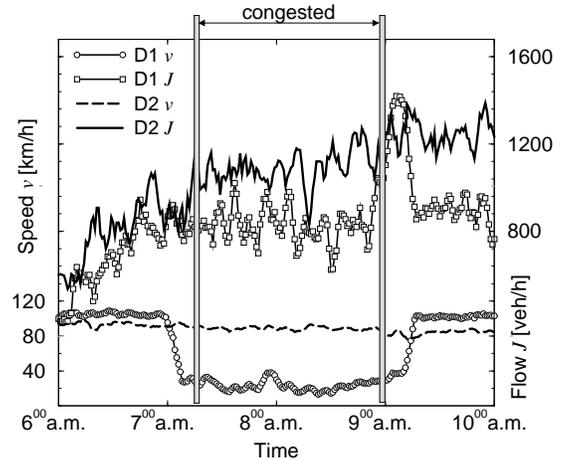,width=0.9\linewidth}
    \caption{Typical time series of flow and velocity. The
      transition from free to congested flow is indicated by a sharp
      fall of the average velocity. Upstream the bottleneck one finds
      a strong reduction of the speed at D1, whereas the flow remains
      nearly constant. Downstream the bottleneck at D2 the outflow
      from a jam is recorded -- the speed is almost constant. It
      increases again after the jam at D1 has dissolved.  For a proper
      characterization of the different states we excluded the
      transition region, e.g.  for the shown realization of a
      congested we restricted our analysis on the part of the
      time-series between the two vertical lines.}
    \label{vser}
  \end{center}
\end{figure}

For a sensible discussion it is plausible to split up the data set
according to the different traffic states. In Fig.~\ref{vser} a
typical time-series of one minute aggregates of the speeds at the
detectors D1 and D2 is shown. The transition from a free-flow to a
congested state is indicated by a sudden drop of the local velocity.
This allows for a undoubted separation of the data set into free-flow
and congested regimes. Then the analysis of the data has been
performed separately for the free-flow and congested states excluding
the transition regime. At D1, the most interesting installation, one
obtains eight different periods of both free-flow and congested
states. These periods are labeled by numbers I through VIII.

Fig.~\ref{vser} also shows the bottleneck-effect given by the
lane-reduction near the intersection. At D1, the cross-section behind
the local defect, one gets a sudden drop in the velocity. On the other
hand, downstream this cross-section one finds only a weak decay of the
velocity which represents the outflow from a jam.

\section{Analysis of single-vehicle data}
\label{sec:single}

In this section the results for the time-headway distributions and
speed-distance relations calculated from single-vehicle data are
presented. For a detailed examination the data set was classified in
two ways. As mentioned above a discrimination between free-flow and
congested states was made, followed by a classification due to local
densities\footnote{In Appendix A it is described how the local density
  is deduced from the data set.}. Whereas the first one was done by a
simple and manual separation by means of the time series of the speed,
the second one requires a more detailed explanation: Every count
belongs to a certain minute, and the local density $\rho$ obtained
during this certain minute is the criterion for the classification.
I.e., it is conceivable that a distance-headway $\Delta x$ is much
larger than the mean distance-headway $\langle\Delta
x\rangle\propto\rho^{-1}$ of the considered period. Moreover, for the
analyses made in this section the traffic states recognized as
stop-and-go traffic are omitted, since the determination of the
related local densities strongly depends on the used
method\footnote{This inevitable behavior of the local density is more
  precisely discussed in Appendix A. Stop-and-go traffic is
  characterized by a large value ($\approx 1$) of the cross-covariance
  between the local density and the local flow as defined by
  (\ref{crosscovdef}) and displayed in Fig.~\ref{crosscov}.}. So only
the synchronized states remain.

\subsection{Time-headway distribution}

\begin{figure}[ht]
  \begin{center}
    \epsfig{file=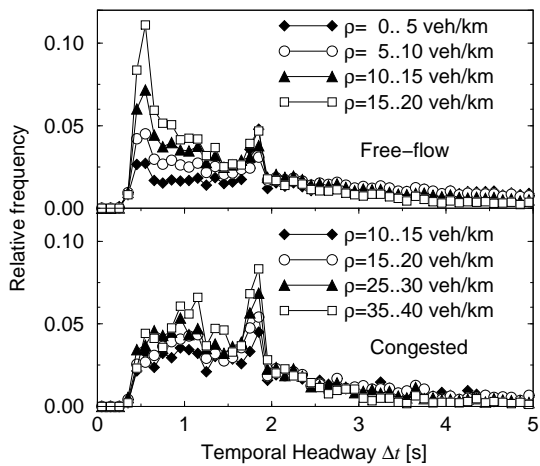,width=0.9\linewidth}
    \caption{Time-headway distribution for different density
      regimes. Top: In free-flow states the $\Delta t$-distribution is
      dominated by two peaks at $0.8\sec$ and $1.8\sec$. Bottom: This
      diagram concerns congested states, the peak at $1.8\sec$
      remains. In both cases there is a significant share of vehicles
      with $\Delta t<1\sec$.}
  \label{thfree}
 \end{center}
\end{figure}

In section \ref{data} the way of calculating the time-headways $\Delta
t$ is described in detail. In principle the accuracy of the
measurement would allow for a very fine resolution of the time-headway
distribution, but in order to obtain a reliable statistics we have
chosen time-intervals of length $0.1\sec$. In Fig.~\ref{thfree} the
time-headway distributions of different traffic states at different
local densities $\rho$ are displayed.
%
%
Regardless of the value of the local density all free-flow
distributions are dominated by a two-peak structure. The first peak at
$\Delta t=0.8\sec$ represents the global maximum of the distribution
and is in the range of time a driver typically needs to react to
external incidents.

On a microscopic level this short time-headways correspond to platoons
of some vehicles traveling very fast -- their drivers are taking the
risk of driving "bumper-to-bumper" with a rather high speed. These
platoons are the reason for the occurrence of high-flow states in free
traffic. The corresponding states exhibit meta stability, i.e. a
perturbation of finite magnitude and duration is able to destroy such
a high-flow state \cite{KernerTGF}. Once such a collapse of the flow
resp. the speed emerges, the free-flow branch can only be reached
again by reducing the local density \cite{KernerTGF,Barlov}. In the
data base considered here such a sharp fall is not observable since
all detected jams are caused by the bottleneck downstream the detector
D1. Additionally, a second peak emerges at $\Delta t=1.8\sec$ which
can be associated with a typical drivers' behavior: It is recommended
and safe to drive with a temporal distance of $\approx 2\sec$
corresponding to a maximum flow of $\approx 1,800\,veh/h$.

Surprisingly, the small time-headways have much less weight in
congested traffic but the peak at $\Delta t=1.8\sec$ is recovered.
Here the background signal is of greater importance. The observed peak
corresponds to the typical temporal headway ($\approx 2\sec$) of two
vehicles leaving a jam consecutively.

However, almost every fourth driver falls below the
$1$-$\sec$-threshold, and this is more likely when the traffic is
free-flowing. Moreover, our results indicate that the small
time-headways are of the highest weight in the transition regime
between free-flow and congested flow.

The common structure of the time-headway distributions in all density
regimes can be summarized as follows: A background signal covers a
wide range of temporal headways, especially for $\tau <1\sec$.
Additionally, at least one peak is to be noticed.

\subsection{Speed distance-headway characteristics}
\label{sec:ov}

Probably the most important information for an adjustment of the speed
is the accessible distance-headway $\Delta x$. This is captured by
several models which use either a stationary fundamental diagram
\cite{KK} or even more directly a so-called optimal-velocity (OV-)
function $v=v(\Delta x)$ as input parameters \cite{Bando94}.
Therefore, a detailed analysis of the speed-distance relationship is
of great importance for the modeling of traffic flow.
%
%
\begin{figure}[ht]
  \begin{center}
    \epsfig{file=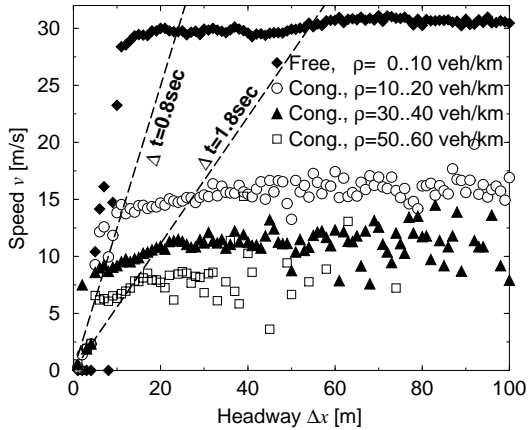,width=0.9\linewidth}
    \caption{The mean speed chosen by a driver depends on both the
      global traffic state and the gap to his predecessor. For a
      better orientation the regions of characteristic temporal
      headways are also displayed.}
    \label{vx}
\end{center}
\end{figure}
By means of Fig.~\ref{vx} it is obvious that the average speed does
not only depend on $\Delta x$ itself, but also on the local density.
In particular, the average speed for large distances in congested
states is significantly lower than for the free flow states, but also
saturated for sufficiently large $\Delta x$.
%
%
\begin{figure}[ht]
  \begin{center} 
        \epsfig{file=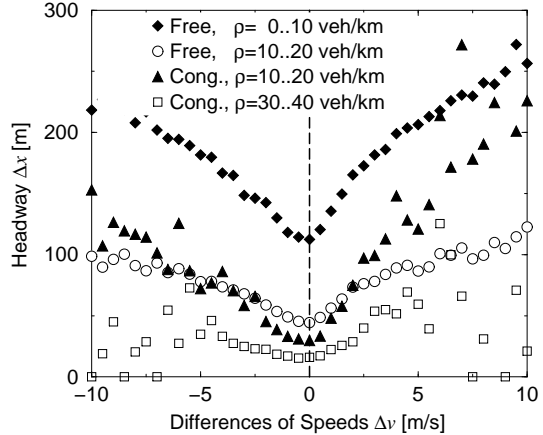,width=0.9\linewidth}
  \caption{The "driving comfort" is not decoupled from more
    "technical" features, i.e. vanishing differences of speed allow to
    drive at the smallest mean distance-headway.}
  \label{xdv}
  \end{center}
\end{figure}
Next we also took into account the velocity differences $\Delta v =
v_n - v_{n+1}$ between consecutive cars ($n$ followed by $n+1$). The
dependence $\Delta x=\Delta x(\Delta v)$ is depicted in
Fig.~\ref{xdv}, where we also discriminated between the interesting
traffic states. The results clearly indicate that $\Delta x$ is
minimized if both cars move with the same velocity, irrespective of
the microscopic state. Note that $\Delta x$ is smaller than the mean
distance derived from the inverse of the local density. Similar
results and comparable conclusions were presented in \cite{Wagner},
where the probability distribution $P(v_t-v_{t+\tau})$ was
investigated. In this context $v_t$ resp. $v_{t+\tau}$ are the speeds
of two arbitrary (not necessarily consecutive) vehicles crossing the
detector with a temporal distance of $\tau$ seconds. They also
observed a peak at $v_t-v_{t+\tau}=0$ for any $\tau$.
%
%
\begin{figure}[ht]
  \begin{center}
    \epsfig{file=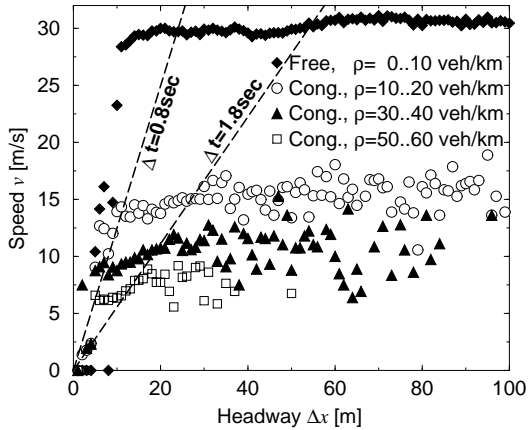,width=0.9\linewidth}
    \caption{If only vehicles with $|\Delta v_n|\leq 0.5 m/\sec$ are
      taken into account, the resulting OV-function differs slightly
      from Fig.~\ref{vx}. Especially in the congested states the
      measurements for $\Delta t\in [0.8\sec , 1.8\sec ]$ are
      proportional to $\Delta x$, beyond $\Delta t=1.8\sec$ the data
      points are rather scattered.}
    \label{vxcomfort}
  \end{center}
\end{figure}

These observations are the motivation to determine an OV-function
using exclusively the data where $|\Delta v_n |\leq 0.5m/\sec$,
because these should be relevant if an empirical OV-function is
demanded as input parameter for traffic models. By using this reduced
data set a better convergence of the OV-function in the high density
regime is observable (Fig.~\ref{vxcomfort}). Nevertheless, at least
the results in the free-flow and congested regime strongly differ.
This indicates that in congested traffic the drivers do not only react
on the distance to next vehicle ahead, they also take into account the
situation at larger distances. It should be noticed that the dropping
of measurements with $|\Delta v_i|>0.5m/\sec$ leaves only a fifth part
of the data, but the quality of the OV-diagrams does not suffer very
much from this restriction.

In order to give explicit measures for the OV-functions in the
different density regimes we used the ansatz:
%
%
\begin{equation}
  \label{ov_function}
  V(\Delta x) = k \left[\tanh(a(\Delta x -b)) + c \right]
\end{equation}
suggested by Bando et al. \cite{Bando94}, where $a,b,c,k$
serve as fit parameters.
%
%
\begin{figure}[ht]
  \begin{center}
    \epsfig{file=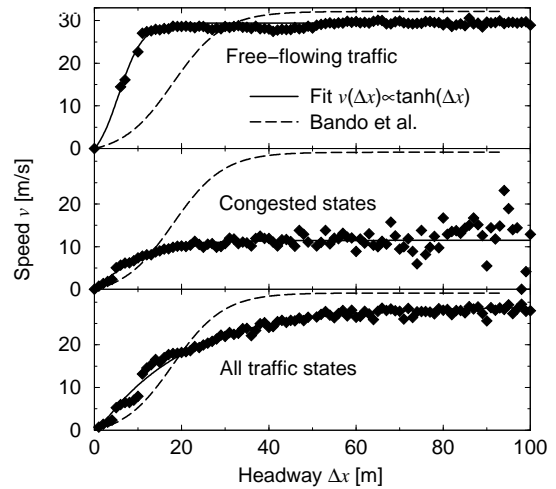,width=0.9\linewidth}
    \caption{Fits of the empirical data using the ansatz
      (\ref{ov_function}). The results are compared with the results
      given in [10]} 
    \label{ovfit}
  \end{center}
\end{figure}
In Fig.~\ref{ovfit} the empirical relations $v=v(\Delta x)$ are
displayed averaging (top) over all states corresponding to free-flow,
(middle) over all congested states and (bottom) over all empirical
data satisfying the restriction $|\Delta v|\leq 0.5m/s$. The
comparison with an empirical OV-function established by analyses of a
car-following experiment on a Japanese highway \cite{Bando94} reveals
a higher value of $v(\infty)$ and a slower increase of the
OV-function.

The characteristic values of the different OV-functions are summarized
in Tab.~\ref{tab:OV_table}, where D denotes the distance where $V(D) =
0.95 \cdot V(\infty)$ holds. The numerical results show that when
averaging over both free-flow and congested states (Fig.~\ref{ovfit},
bottom) the asymptotic regime of the OV-function is reached at much
larger distances.
%
%
\begin{table}[h]
  \begin{center}
    \leavevmode
    \begin{tabular}[h]{l||c|l}
      & D & $V(\infty)$ \\
      \hline
      Bando \cite{Bando94} & 42.39  & 32.14 \\
      Free-flow (top)& 14.11 & 29.43 \\
      Congested (middle) & 26.70 & 11.47\\
      All States (bottom) & 57.70 & 28.64\\
    \end{tabular}
    \caption{Characteristic parameters of the OV-functions. The
      asymptotic values of the velocity $V(\infty)$ as well as the
      distance D where $95\%$ of $V(D)$ are exceeded (refer to
      Fig.~\ref{ovfit}).}
    \label{tab:OV_table}
  \end{center}
\end{table}

Our results for the OV-functions can be summarized as follows. In the
free-flow regime the function are characterized by a steep increase at
small distances corresponding to the small time-headways discussed in
the previous subsection. For synchronized states it is remarkable that
the asymptotic velocity takes a rather small value. Furthermore, our
results show that it is necessary to distinguish between the traffic
states in order to get a more precise description of the speed-headway
relation.

\section{The fundamental diagram}
\label{sec:fd}

In this section we present results on the fundamental diagram based on
time-averaged data. The present data set allows for a free choice of
the averaging interval and overlaps. Here we compare the results
obtained for one- and five-minute intervals. At the end of this
section we discuss different methods in order to establish the
stationary fundamental diagram.
%
%
\begin{figure}[ht]
  \begin{center}
    \epsfig{file=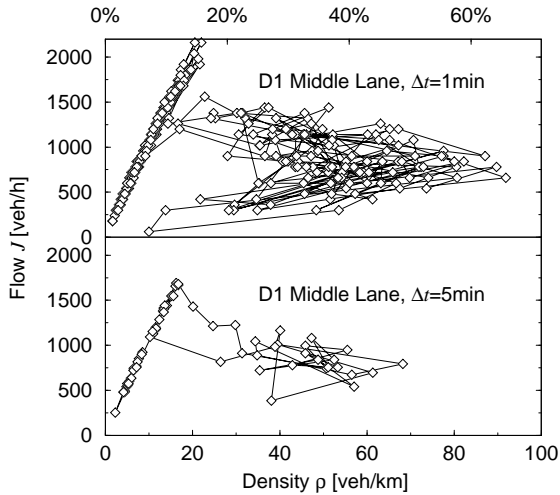,width=0.9\linewidth}
    \caption{Fundamental diagrams for different averaging intervals
      $\Delta t$. The upper diagram shows time-traced data averaged
      over $\Delta t=1\,min$ while for the lower diagram $\Delta
      t=5\,min$ has been used (for an explanation of the structure see
      also Appendix A and Fig.~\ref{simu}). The relative occupation is
      calculated using the maximal density during the measuring
      period, $\rho_{max}=140\,veh/km$. }
    \label{fund_scat}
  \end{center}
\end{figure}
In Fig.~\ref{fund_scat} fundamental diagrams for averaging intervals
$\Delta t$ of one and five minutes are shown. Beyond the trivial
effect that longer averaging intervals lead to a reduction of the
fluctuations, one observes that both the extremal values of the
density and the flow decrease with growing $\Delta t$. Moreover, the
small flow values at very low densities are averaged out if five
minute intervals are chosen. One might ask whether longer $\Delta t$'s
hide some real structure of traffic states or whether the additional
structure in the one minute intervals is a statistical artifact. From
our point of view the results for the low density branch, which agree
for both averaging intervals, indicate that a one-minute interval is
sufficient to establish the systematic density dependence of the flow.
Beyond that microscopic states with short lifetimes can be detected
using short time-intervals which makes the one-minute intervals
preferable.

The uncommon structure structure of the flow-density relation for
small speeds, especially the return into the origin of the coordinate
system, must be traced back to the method of the determination the
local density via $J/v$, since the occupation itself was not
accessible in the underlying data set. This behavior will be explained
in more detail in Appendix A.
%
%
\begin{figure}[ht]
  \begin{center}
    \epsfig{file=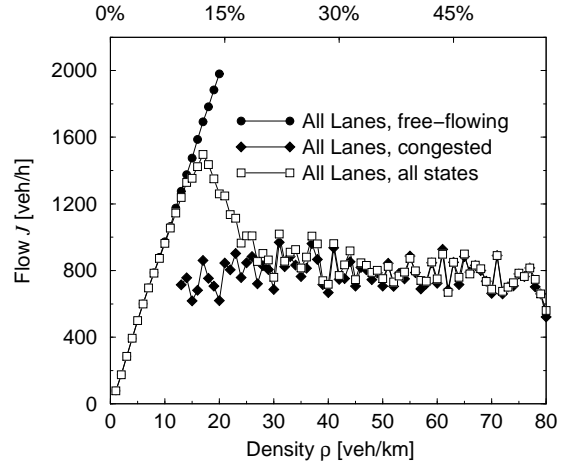,width=0.9\linewidth}
    \caption{Mean flow-density relation using the complete data
      set. The continuous curve corresponds to an average of all flow
      values for a given density while the discontinuous line is
      obtained discriminating between free-flow and congested states.
      Using the latter procedure a non-unique behavior of the
      flow-density relation is observable which can also be seen near
      local defects in driven systems. Data points of the congested
      branch have been dropped for low densities due to the unreliable
      statistics.}
    \label{fund_av}
  \end{center}
\end{figure}

In order to obtain the {\em stationary} flow-density relationship we
generated histograms from the fundamental diagram. Due to the problems
of the density estimation in stop-and-go traffic we omit these states
in the further discussion\footnote{See the following section for the
  identification of stop-and-go states.}. In Fig.~\ref{fund_av} the
results for two averaging procedures are displayed.  The continuous
form of the fundamental diagram has been obtained by averaging over
all flow values of a given density, while the discontinuous shape has
been obtained discriminating between free-flow and congested traffic.
It should be mentioned that the shape of the continuous stationary
fundamental diagram also depends on the statistical weight of
free-flow and congested states.  Therefore from our point of view it
is necessary to distinguish between the different states in order to
obtain reasonable results for the stationary fundamental diagram.

Using the latter method it turns out that for high densities the
average flow takes a constant value in a wide range of density. This
plateau formation is similar to what is found in driven systems with
so-called impurity sites or defects, where in a certain density regime
the flow is limited by the capacity of the local defect
\cite{Janowsky92,Schutz93,Csahok94,Chowdhury98}. Here the bottleneck
effect is produced by lane-reduction as well as by the large activity
of the on- and off-ramps at the intersection (see Section \ref{data}).

\section{Time series analysis} 
\label{sec:time}

As already mentioned in the introduction we propose {\em objective}
criteria for the classification of different traffic states using
standard methods of time-series analysis. Beyond that we will show
that these methods allow for a further characterization of the
different states.

\subsection{Auto-covariance function}
\label{sec:aclocal}

The first quantity to consider is the auto-covariance
%
%
\begin{equation}
  \label{autocovdef}
 ac_x(\tau) = \frac{\langle x(t)x(t+\tau) \rangle -\langle x(t)\rangle^2} {\langle x^2(t) \rangle -\langle x(t)\rangle^2}
\end{equation}
of the aggregated quantities $x(t)$. The brackets $\langle \dots
\rangle$ indicate the average over a complete period of a free-flow or
congested state.
%
%
\begin{figure}[ht]
  \begin{center}
    \epsfig{file=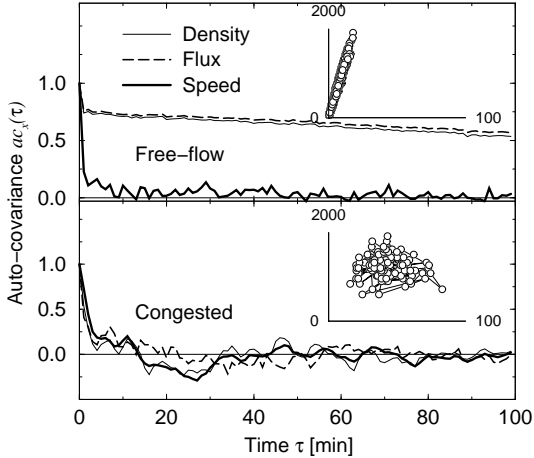,width=0.9\linewidth}
    \caption{Typical auto-covariance functions of the local density, the
      average velocity and the flow in a free-flow (top) and congested
      (bottom) state. The insets show the fundamental diagram
      corresponding to the chosen time-interval.}
    \label{autocov}
  \end{center}
\end{figure}
In Fig.~\ref{autocov} the auto-covariances of one-minute aggregates of
the density, flow and average velocity of a free-flow and a congested
state are shown. In the free-flow state the average speeds are only
correlated on short time scales whereas long-ranged correlations are
present in the time series of local density as well as of the flow.
This implies that no systematic deviations of the average velocity
from the constant average value are observable, while the density and
therefore also the flow vary systematically on much longer time scales
up to the order of magnitude of hours.

This behavior of the auto-covariance is clearly contrasted with the
behavior found in synchronized traffic, where {\em all} temporal
correlations are short-ranged irrespective of the chosen observable.
Both results show that longer time scales are only apparent in slow
variations of the density during a day, while the other time-series
reveal a noisy behavior.

\begin{figure}[ht]
  \begin{center}
    \epsfig{file=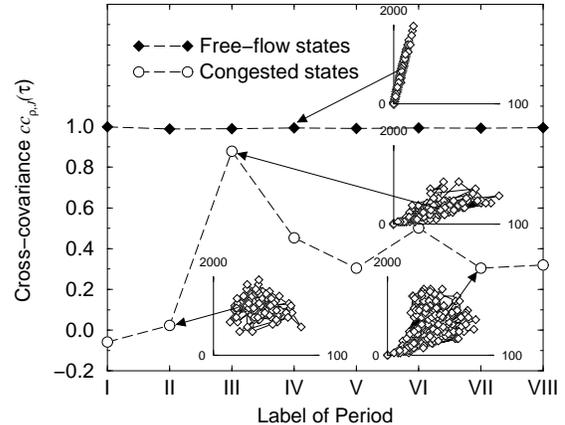,width=0.9\linewidth}
    \caption{Strong correlations between density and flow, indicated by
      $cc_{\rho,J}(0) \approx 1$, can be found in both free-flow and
      stop-and-go traffic. By contrast synchronized states are
      characterized by weak correlations between density and flow.
      Congested states where transitions between synchronized and
      stop-and-go traffic appear lead to intermediate values of
      $cc_{\rho,J}(0) \approx 0.2\ldots 0.5$. The different periods of
      free-flow and congested traffic are labeled by I through VIII
      (see also Section~\ref{data}). The dashed lines are a guide to
      the eyes only.}
    \label{crosscov}
  \end{center}
\end{figure}

Furthermore, the cross-covariance
%
%
\begin{equation}
  \label{crosscovdef}
 cc_{x,y}(\tau) = \frac{\langle x(t)y(t+\tau) \rangle -\langle x(t)\rangle\langle y(t+\tau)\rangle} {\sqrt{\langle x^2(t) \rangle -\langle x(t)\rangle^2}\sqrt{\langle y^2(t) \rangle -\langle y(t)\rangle^2}}
\end{equation}
indicates the strong coupling between flow and density in the
free-flow regime (Fig.~\ref{crosscov}). This implies that the
variations of the flow are mainly controlled by density fluctuations
while the average velocity is almost constant. Again the results for
synchronized states differ strongly. Here all combinations of flow,
density and average velocity lead to small values of the
cross-covariance also supporting the existence of irregular patterns
in the fundamental diagram.
%
%
Therefore the covariance analysis of the empirical data are in
agreement with the interpretation of empirical data given in
\cite{Kerner961,Kerner972}, where synchronized states first have been
identified. The synchronized states can be distinguished from
stop-and-go traffic \cite{Kerner981} using the same methods.
%
%
\begin{figure}[ht]
  \begin{center}
    \epsfig{file=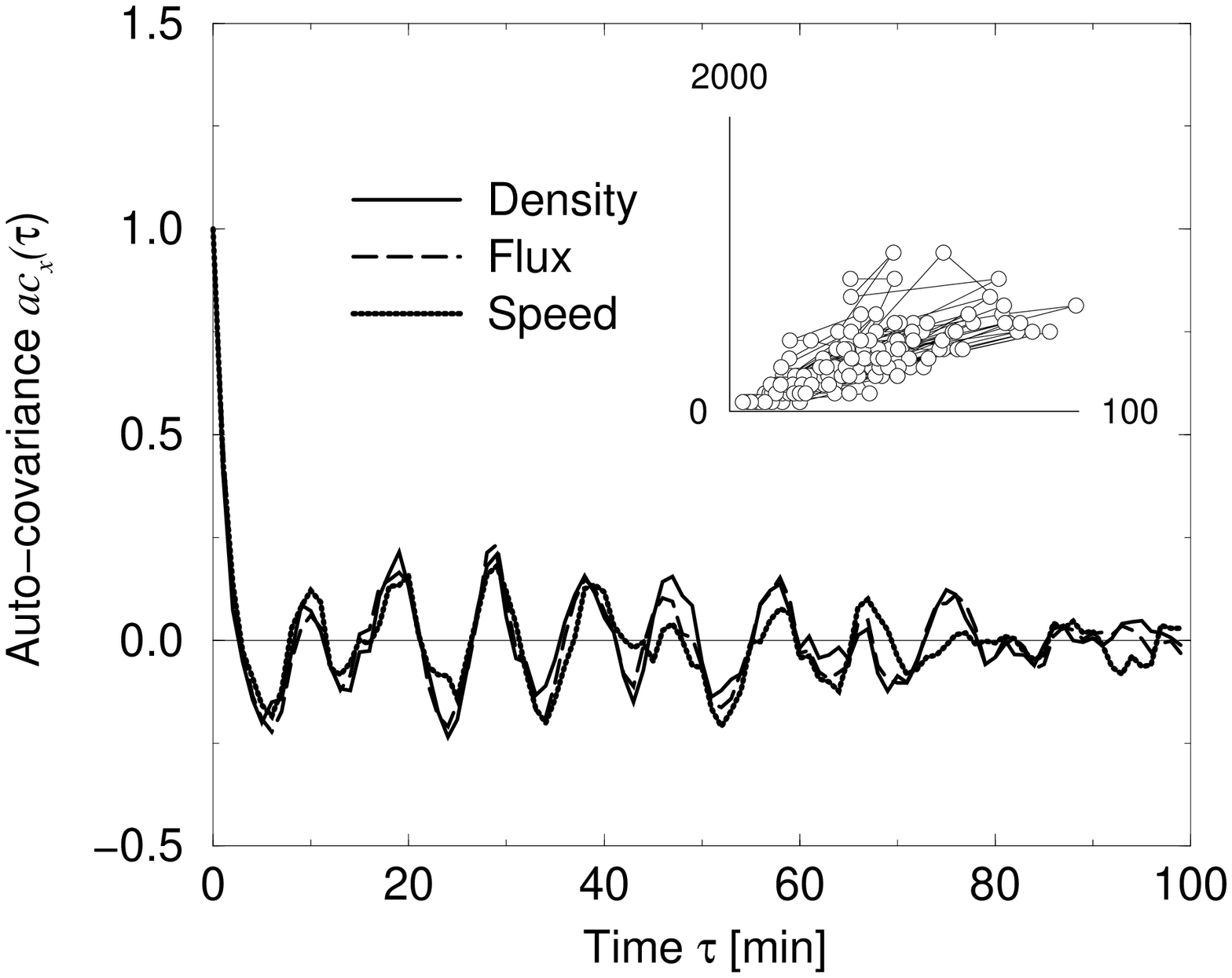,width=0.9\linewidth}
    \caption{The auto-covariances of all three local
      measures in stop-and-go traffic. All three local measures show
      oscillations around $0$ with a period of $\approx 10\,min$.  }
    \label{cong2autocov}
  \end{center}
\end{figure}
Similar to free-flow states stop-and-go traffic is characterized by
strong correlations between density and flow ($cc_{\rho,J}(0) \approx
1$). Beyond that also the auto-covariance function shows an
interesting behavior namely an oscillating structure for all three
quantities of interest. The period of these oscillations is given by
$\approx 10\,min$. This result is in accordance with measurements by
K\"uhne \cite{Kuehne}, who found oscillating structures in stop-and-go
traffic with similar periods.

\subsection{Transitions between the different states}

\begin{figure}[ht]
  \begin{center}
    \epsfig{file=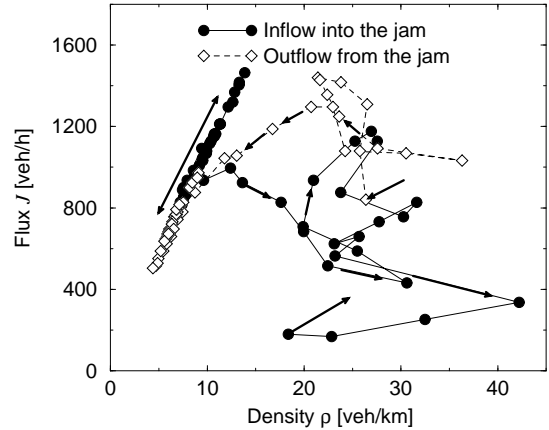,width=0.9\linewidth}
    \epsfig{file=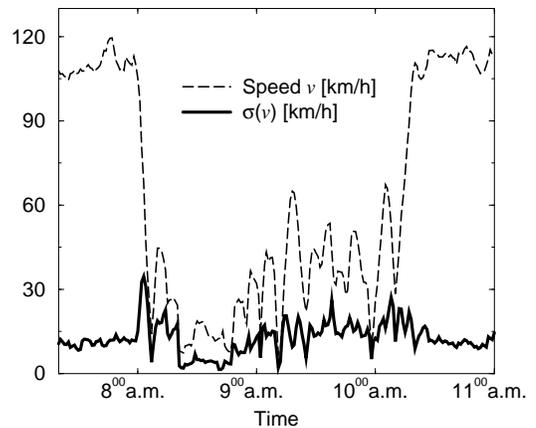,width=0.9\linewidth}
    \caption{Transitions between free-flow and congested traffic. The
      upper diagram shows one minute aggregates of the flow near a
      transition from free-flow to a congested state and vice versa.
      For comparison the time-series of the one-minute aggregates of
      the local speed and their standard deviations are given in the
      diagram below.}
   \label{fdtrans}
  \end{center}
\end{figure}

These previous results show that the time-series analysis allows for
an identification of different traffic states. Now we focus on the
transition regime. Compared to the typical life-time of a free-flow or
a congested state the transition is of short duration of approximately
the order of magnitude of fifteen minutes (see Fig.~\ref{fdtrans} for
a typical time-series of the local speed including a congested state).

%
%

Transitions from free-flow to both congested states are observable in
the data set. The transitions take place at densities significantly
lower than the density of maximum flow, since the transitions are
initiated by a reduction of the capacity of the bottleneck and not by
a continuous increase of the local density.  We also want to mention
that the congested states often are composed of stop-and-go and
synchronized states, i.e. during a time-series corresponding to
congested traffic frequently transitions between both congested states
occur.

The results of several other empirical investigations suggest that the
transition between free and congested flow is accompanied by a peak of
the velocity variance at the transition
\cite{Helbi,Kuehne,Helbing971}. Our analysis clearly does not support
this result. The existence and height of the peak is closely related
to the length of the averaging intervals. But of course these peaks
are numerical artifacts. They show up because the time interval
includes two different states but do not reflect any further
characteristics of the transition.

\subsection{Correlation between different lanes}

In addition to the irregular pattern in the fundamental diagram it has
been argued that a characteristic feature of the synchronized states
is the strong coupling between different lanes
\cite{Kerner961,Kerner972}. These interpretations are mainly based on
the fact that the average speeds on the different lanes approach each
other.  Here this effect is not observable because due to the
speed-limit even in the free-flow regime the average velocities on
different lanes only slightly differ (the average speed in the
free-flow regime on the left lane is given by $\approx 120\,km/h$ and
on the two other lanes by $\approx 100\,km/h$).
%
%
\begin{figure}[ht]
  \begin{center}
    \epsfig{file=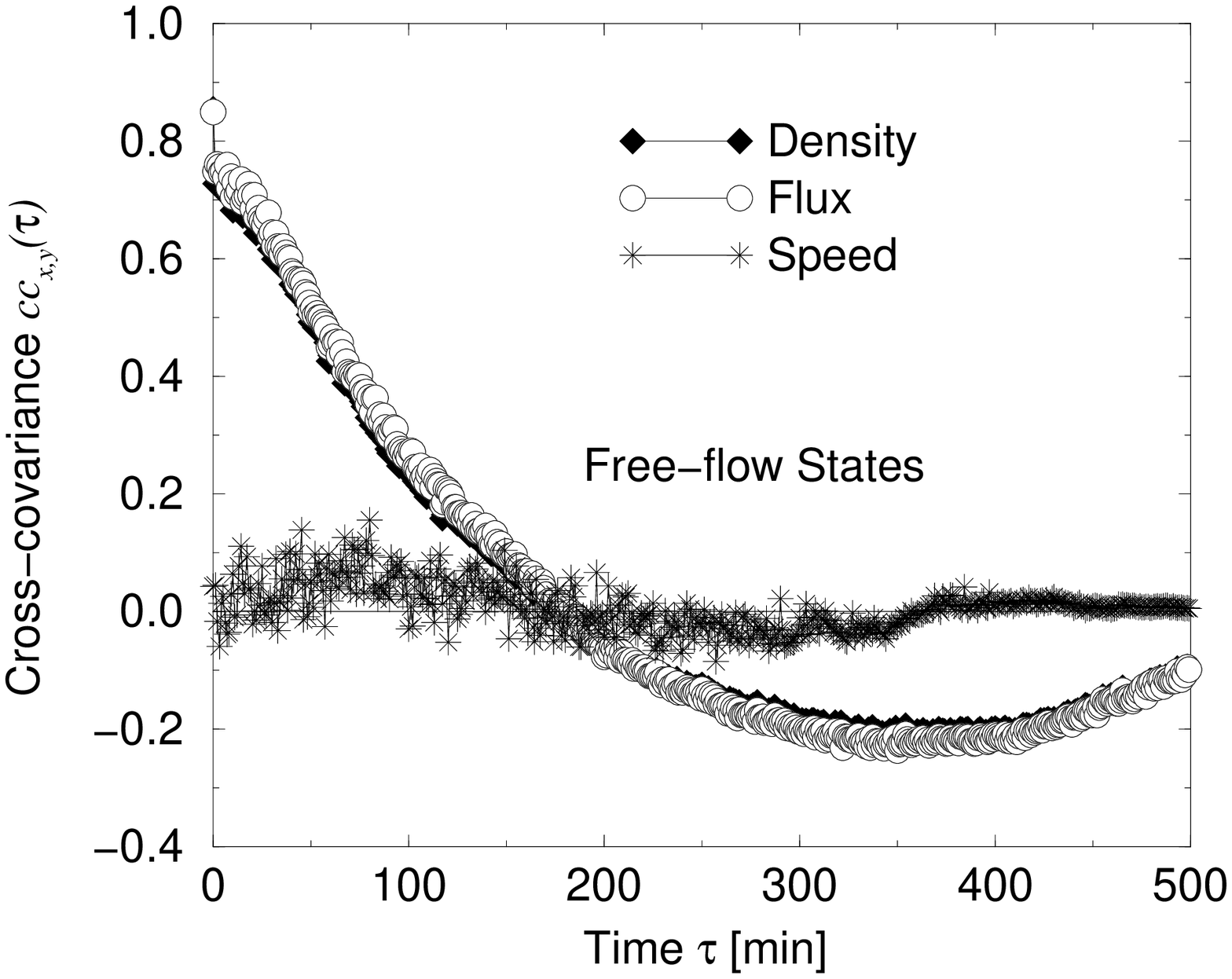,width=0.9\linewidth}
    \epsfig{file=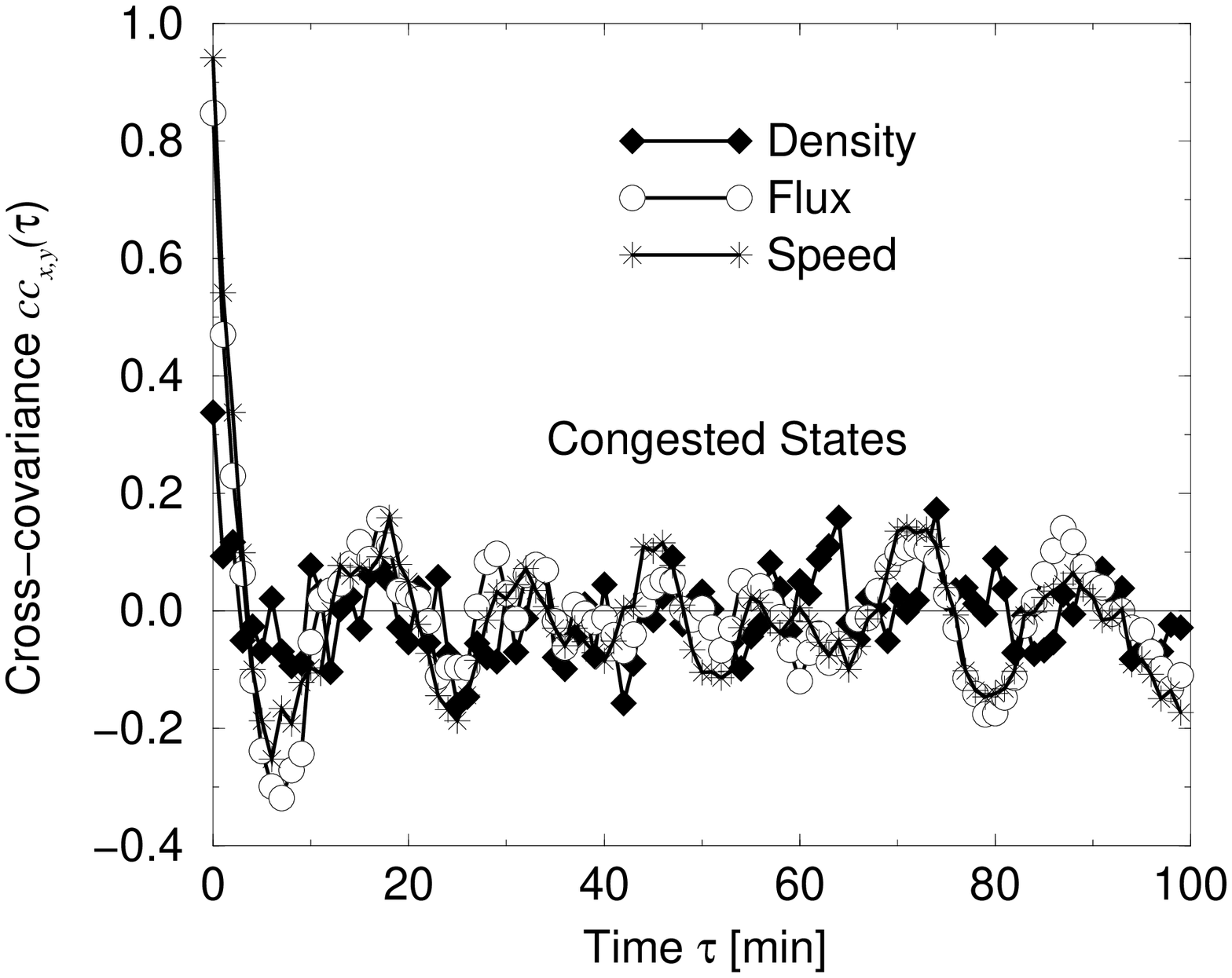,width=0.9\linewidth}
    \caption{Cross-covariance of flow, density and speed of
      different lanes in the free-flow regime compared to congested
      states.}
    \label{lanecc}
  \end{center}
\end{figure}
Therefore we calculated $cc_{x_i,x_j}(\tau)$ in order to quantify this
coupling effect. Here $x_i$ denotes the flow, density, or speed on
lane $i$. In Fig.~\ref{lanecc} the cross-covariances of different
lanes belonging to the same driving direction are shown.  The coupling
between flow and density in the synchronized state is comparable to
the free-flow state. It is also apparent that the free-flow signal is
veritable on long time-scales, while in synchronized states the
correlations rapidly decay with time. Again this result mainly
reflects the daily variation of the density.

The synchronization of the different lanes is indicated by large value
($cc_{v_i,v_j}(0) \approx 0.9$) of the cross-covariance of the speed
at $\tau = 0$, while the time series of the speed on different lanes
in free-flow are completely decoupled.
   
\subsection{Time-series of the single-vehicle data}
\label{sec:acsingle}

In the previous section it could be shown that the relevant
time-scales are identifiable by time-series analyses.  Here these
methods, in particular a generalization of the cross-covariance
function, will be used in order gain further information on the
different microscopic states.

By using the single-vehicle data directly it is not possible to
evaluate the time-dependence of $ac_x(t)$ in realistic units because
the time-intervals between consecutive signals strongly fluctuate.
Instead of the temporal difference $\tau$ now the number of cars $n$
passing the detector between vehicle $i$ and $j = i+n$ is used.
%
%
\begin{figure}[ht]
  \begin{center}
    \epsfig{file=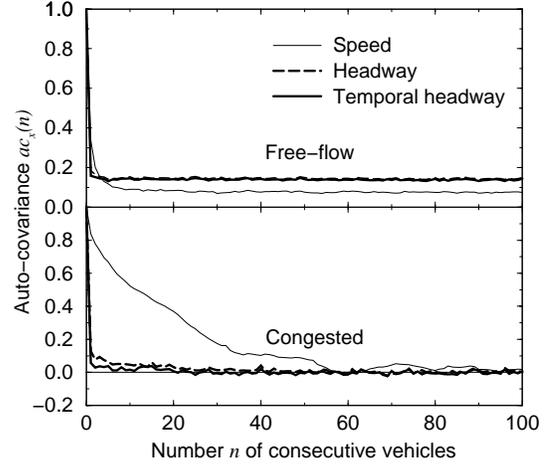 ,width=0.9\linewidth}
    \caption{A long-range signal is obtained from the
      auto-covariance of spatial and temporal headways among
      free-flowing vehicles (top). The diagram below is related to
      congested states and makes clear that the correlations of the
      speeds are large on short scales (up to $50$ vehicles), whereas
      that of the spatial and temporal headway decays very abruptly.}
    \label{svtime}
  \end{center}
\end{figure}

The behavior of the auto-covariance in the free-flow regime can be
characterized as follows. For small $n$'s one observes a steep
decrease of $ac_x(n)$ while asymptotically slow decrease is found. The
crossover from a fast to a slow decay has been observed for a small
number of cars ($n \approx 5$).

In the free-flow regime the single vehicle data basically support the
results obtained for the aggregated data, namely a strong coupling
between the temporal headways (the single-vehicle analogy of the flow:
$J\propto\Delta t^{-1}$) and the distances (corresponding to the
density: $\rho\propto\Delta x$). Moreover, the slow asymptotic decay
is mainly due to the daily variation of the density. A different
behavior has been found for $ac_v(n)$.  First the decay for small $n$
is not as fast as for the other signals and second the function decays
faster asymptotically. The asymptotic behavior, in turn, is in
accordance with the result drawn from aggregated data. But from our
point of view the slower decrease for short distances is of special
interest. It implies that also in the free flow regime small platoons
of few cars moving with the same speed are formed. These platoons lead
to the peak at $\Delta t = 0.8\sec $ in the time-headway distribution.

Having Fig.~\ref{svtime} in mind $ac_x(n)$ of congested-state
quantities behaves similarly, except for two differences: First of all
no long-ranged signal is present for all quantities of interest and
second the decay of $ac_{v}(n)$ for small $n$ is much weaker than in
the free-flow regime. This leads to the following picture of the
microscopic states in synchronized flow: Similar to the-free flow
regime platoons are formed with cars moving at the same speed, but in
synchronized flow these platoons are much larger (of the order of
magnitude of some ten vehicles).

\section{Summary and conclusion}

In this paper a detailed statistical analysis of single-vehicle data
of highway traffic is presented. The data allow to analyze the
microscopic structure of different traffic states as well as a
discussion of time-averaged data.

Using the single-vehicle data directly we calculated the time-headway
distribution and the headway-dependence of the velocity. Both
quantities are of great interest for modeling of traffic flow, because
they can be directly compared with simulation results
\cite{Chowdhury97} or are even used as input parameters \cite{Bando94}
for several models.

Our analysis of the time-headway distribution has revealed a
qualitative difference between free-flow and synchronized states.  The
time-headway distribution of free-flow states shows a two-peak
structure. The first peak is located at very small time-headways
($\Delta t\approx 0.8\sec$) while a second peak shows up at $\Delta
t\approx 1.8\sec$. The second peak is also observed in congested flow
but smaller time-headways are significantly reduced.  The small
time-headways correspond to very large values of the flow.  Therefore
the peak at small time-headways can be interpreted as a microscopic
verification of meta-stable free-flow states.
 
Similar results have also been obtained for the speed distance
relation, the so-called OV-function \cite{Bando94}. It also turned out
that it is necessary to distinguish between free-flow and congested
states. In particular the asymptotic velocities in free-flow and
congested states differ strongly. Moreover a global average leads to
different characteristics at small distances.

For comparison with earlier empirical investigations
\cite{Kerner961,Kerner972,Kerner981,Helbing971} we also have used
aggregated data in order to calculate the fundamental diagram.  Our
results indicate that one minute intervals are preferable compared to
five minute periods (see \cite{Helbi} for comparison) although this
short intervals lead to larger fluctuations. Nevertheless, from our
point of view, the single-vehicle data suggest that these fluctuations
are not an artifact of the short averaging procedure but represent the
complex structure of the different traffic states.

The data have also been used to calculate a stationary fundamental
diagram.  Again our results show that it is necessary to distinguish
between free-flow and congested states in order to get reliable
results for the average flow at a given density. Then one obtains a
discontinuous form of the fundamental diagram and a non-unique
behavior of the flow at low densities.

Using the auto-co\-va\-ri\-ance function and cross-co\-va\-ri\-ance
for the different time-series we were able to identify three
qualitatively different microscopic states of traffic flow, namely the
free-flow, synchronized and stop-and-go traffic \cite{KernerTGF}. The
free-flow states are characterized by a strong coupling of the flow
and density and beyond that by a slow decay of the related
auto-covariance functions. This implies that as far as a free-flow
state is present the flow solely depends on the density. The
time-scale which governs the asymptotic decay of the auto-covariance
function is also determined by the daily variance of the density.

As shown in section~\ref{data} one can easily distinguish free-flow
and congested-states. By contrast it is much more difficult to
separate between time-series belonging to stop-and-go and synchronized
states by inspection. Therefore an {\em objective} criterion is of
great interest. It turns that the time-series analysis provides such
criterion. Synchronized states are indicated by small values of the
cross-covariance between flow, speed and density.  Moreover the
auto-covariance function is short-ranged for all three quantities.
These results reflect the completely irregular pattern in the flow
density plane \cite{Kerner961,Kerner972} found for synchronized
states.  By contrast, in stop-and-go traffic flow and density are
strongly correlated. On the other hand, the auto-covariance function
reveals an oscillating structure \cite{Kuehne} with a period of the
order of ten minutes.  In addition it was found that transitions
between free-flow and congested flow are rare but transitions between
the different congested states are more frequent.

The auto-covariance functions of the single-vehicle data have suggest
that in the free-flow regime as well as in synchronized states
platoons of cars moving with the same velocity can be observed.
Presumably the platoons in the free-flow regime lead to the peak at
$\Delta t \approx 0.8$ in the time-headway distribution and therefore
to very large values of the flow.

From our point of view our results have important implications for the
theoretical description of traffic flow phenomena. The short
distance-headways present in free-flow traffic are only possible when
drivers anticipate the behavior of the vehicles in front of him
\cite{Knospe}. Anticipation is less important in congested traffic.
Another important effect is reflected by the gap-dependence of the
velocity at high densities. Here we observe a small asymptotic
velocity. This implies that drivers tend to hold their speed in dense
states, another feature which has to be captured by traffic models.
Finally, one has to take into account the reduced outflow from a jam
which has been verified by other authors and is supported by our
results.

In conclusion the analysis of single-vehicle data leads to a much
better understanding of the microscopic structure of different traffic
states. Although our results give a consistent picture of the
experimental facts on highway traffic an enlarged data set or data
from other detector locations would be very helpful in order to settle
the experimental findings. First of all a series of counting loops
would allow a more detailed analysis of the spatio-temporal structure
of highway traffic, additional data from on- and off-ramps would help
to discriminate between bulk and boundary effects.

\section*{Appendix A: }

In principle one could use the single vehicle data directly in order
to establish the velocity-flow relationship because the speed and the
time-headway (which is proportional to the inverse flow) of individual
cars are provided by the detector.  Unfortunately, an interpretation
of these results is difficult because of the extreme fluctuations of
the experimental data. Therefore we used aggregated data in order to
determine a fundamental diagram. In particular we show the
flow-density relationship of one and five minute aggregates.

While the local flow is directly given in the data set one has to
calculate the temporally averaged local densities $\rho$ at the
detector because the coverage\footnote{The coverage of a detector
  denotes the fraction of time when the detector is occupied by
  vehicles.} of a detector is not provided here. The local density can
be calculated via the relation
%
%
\begin{equation}
  \label{eq:densitydef}
  \rho  = J/ v,
\end{equation}
where $J\propto N$ is closely related to the total number of cars $N$
crossing the detector during the time interval $\left [ t,t+\Delta t
\right ]$, and $v=\sum v_n(t)/N$ the average velocity of the cars.
Note that both the velocity $v_n(t)$ of the individual cars and the
flow $J$ are directly accessible. Therefore this method should give
the best estimate for the local density $\rho$ as long as the velocity
$v_n(t)$ represents a characteristic value of the local speed.

Problems using this kind of density calculation may arise from the
strong fluctuations of the speed, especially in stop-and-go traffic.
Then the velocity recorded by the detector gives a measure of the
typical velocity of {\em moving} vehicles while the periods when cars
do not move are not taken into account.

%
%
\begin{figure}[ht]
  \begin{center}
    \epsfig{file=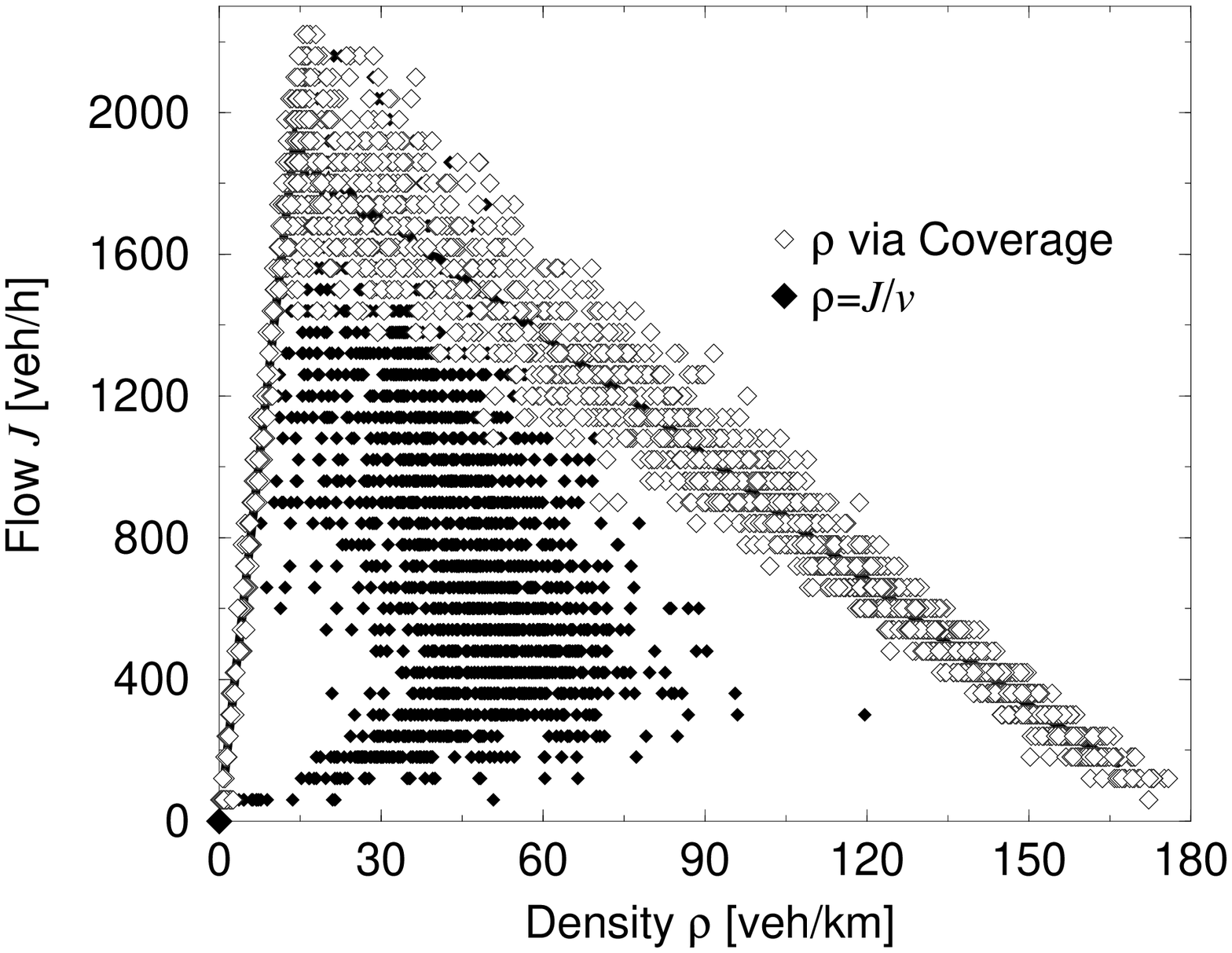,width=0.85\linewidth}
    \epsfig{file=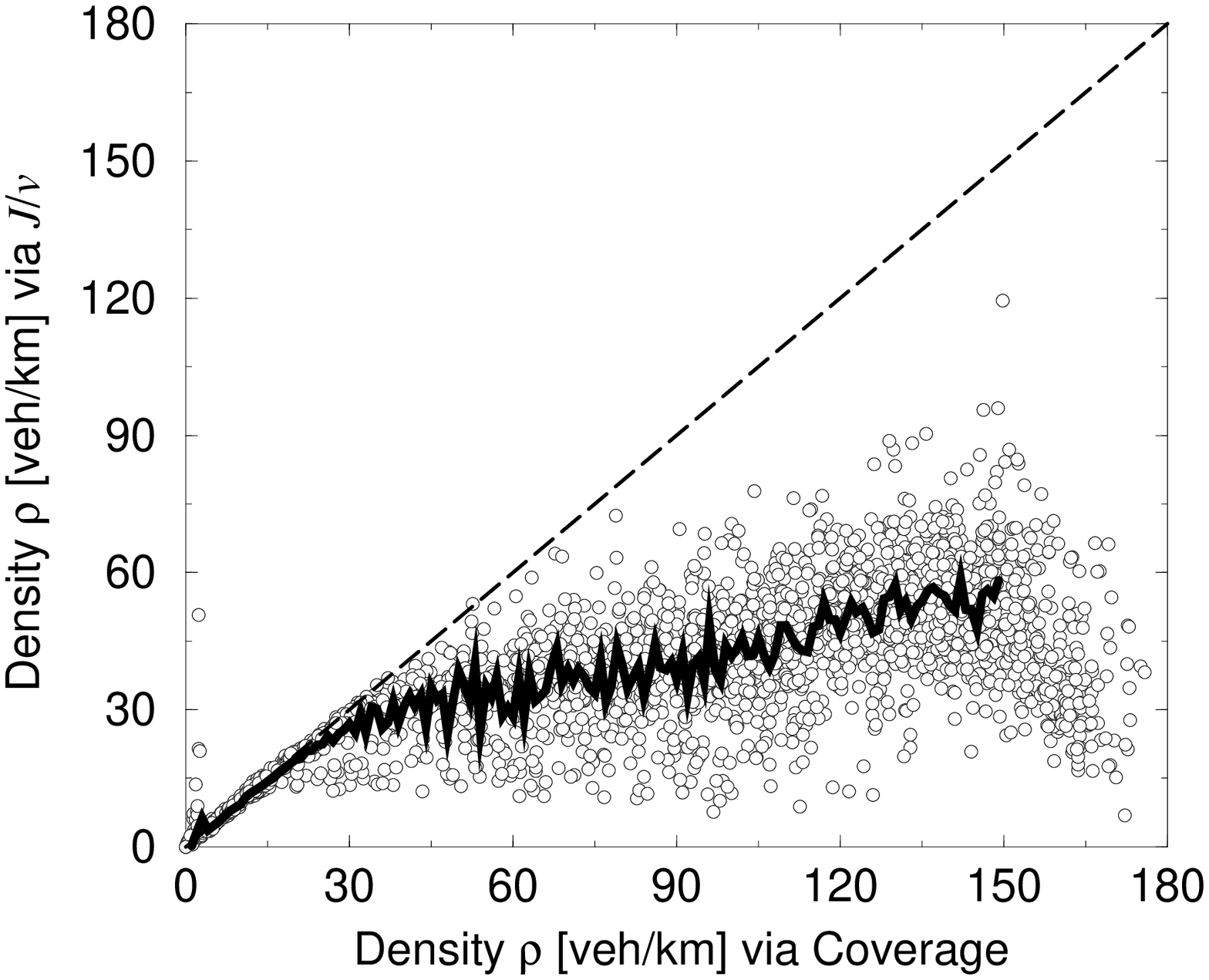,width=0.85\linewidth}
    \caption{Simulation results using different methods for the calculation
      of the local density. Top: The filled symbols correspond to an
      estimate calculated using (\ref{eq:densitydef}), while the open
      symbols represent the coverage of the detector. Bottom:
      Comparison of both estimates. For $\rho>30~veh/km$ both methods
      strongly deviate from each other. Note that the density where
      these differences occur strongly depend on the chosen
      calibration of the model.}
    \label{simu}
  \end{center}
\end{figure}

Fig.~\ref{simu} illustrates the effect of the different measuring
procedures using computer simulations. During the simulation of a
continuous version of the NaSch-model (see Appendix B for a definition
of the model) we used two kinds of detectors. The first detector is
located at a link between two lattice sites. At this link we perform
measurements of the number of passing cars (i.e. the local flow) and
their velocity. Then the local density is calculated via
(\ref{eq:densitydef}). This result is compared with direct
measurements of the local density where the average occupation on a
short section of the lattice is detected. The figure shows that both
estimates of the local density are in good agreement at low densities
while at large densities the estimates may strongly differ.  The
different estimates for the local density lead to different shapes of
the fundamental diagram. Estimating the density via the occupation of
the detector we get the well known form of a high density branch while
the calculation of the local density via (\ref{eq:densitydef}) leads
to a pattern which is similar to free-flow states but with a much
smaller average velocity.

Similar patterns have also been found in our data set (see
Fig.~\ref{fund_scat}). Therefore the simulation results indicate that
these periods correspond to stop-and-go traffic. Data points
representing blocked cars are located in the origin of the fundamental
diagram. Using a coverage-based density the points belonging to the
same period would be shifted to the right. A deadlock situation would
approach ($\rho_{max}$,0), with $\rho_{max}$ the maximum density. We
suppose that $\rho_{max}=140\,veh/km$.

Finally we want to mention that this problem cannot be circumvented
using the speed-flow relation because one is still left with the
problem of overestimating local speeds in stop-and-go traffic.

\section*{Appendix B:}

The simulation results have been obtained using a space-continuous
version of the Nagel-Schreckenberg (NaSch) model \cite{Nagel92} for
single-lane traffic. Analogous to the NaSch model the velocity of the
$n$-th car in the next time step is determined via the following four
rules which are applied synchronously to all cars:
 
\noindent {\it Step 1: Acceleration}\\
\phantom{xxxx}$V_n \rightarrow min(V_n+1,V_{max})$

\noindent{\it Step 2: Deceleration (due to other vehicles)}\\ 
\phantom{xxxx}$V_n \rightarrow min(V_n,d_n)$.

\noindent{\it Step 3: Randomization}\\
\phantom{xxxx}$V_n \rightarrow max(V_n-rand(),0)$ with $rand()\in [0,1]$ 

\noindent{\it Step 4: Movement}\\
\phantom{xxxx}$X_n \rightarrow  X_n + V_n$. 

The velocity of the n-th car $V_n$ is given in units of $5 m/s$.
$V_{max}$ denotes the maximum velocity, $X_n$ the position of the
cars, $d_n= X_{n+1}-X_n-1$ the distance to the next car ahead. $X_n$
and $d_n$ are also given in units of $5m$ (the length of the cars).
$rand()$ is a random number between 0 and 1.  In our simulation we use
$V_{max} = 8$ which corresponds to $40~m/s$ in realistic units.

The discrete NaSch model is also able to generate such fundamental
diagrams, but with a worse resolution -- the line of stop-and-go
traffic has a rather steep slope. This is why we decided to use the
continuous version of the NaSch model. Note that beside the better
spatial resolution of the continuous version, no qualitative
difference between the continuous and discrete version of the model
have been found.

\vspace{0.5cm}

\noindent{\bf Acknowledgments:} It is our pleasure to thank 
B.S. Ker\-ner and D. Chowdhury for fruitful discussions. The authors
are grateful to "Landschaftsverband Rheinland" (Co\-log\-ne) for data
support, to "Systemberatung Povse" (Herzogenrath) for technical
assistance, to the Ministry of Economy, Technology and Traffic of
North-Rhine Westfalia, and to the German Ministry of Education and
Research for the financial support within the BMBF project "SANDY".

\end{document}